# A Survey and Insights on Deployments of the Connected and Autonomous Vehicles in US

Sanchu Han
ABC Technologies Inc.
Sanchu.han@futurewei.com

**Executive Summary**
CV/ITS (Connected Vehicle, Intelligent Transportation System) and AV/ADS (Autonomous Vehicle, Automated Driving System) have been emerging for the sake of saving people lives, improving traffic efficiency and helping the environment for decades. There are separate efforts led respectively by USDOT with state DOTs for CV, and private sectors through market driven approach from start-ups and technology companies for AV.  By CV/ITS effort there are 97 deployments of V2X communications utilizing the 5.9 GHz band, 18,877 vehicles with aftermarket V2X communications devices, and 8,098 infrastructure V2X devices installed at the roadsides. However, CV/ITS still cannot be massively deployed in US markets due to lack of regulations, dedicated wireless spectrum bands, sustainable financial & business models with mature supply chain, etc. In the other side, technology-driven AV market has been much slower than expected mainly because of immaturity of AI technology to handle different complex driving scenarios in a cost effective way. In this paper, we first present these two parallel journeys focusing on the deployments including operating models, scenarios and applications, evaluations and lessons learning. Then, come up with recommendations to a cooperative CAV approach driving a more feasible, safer, affordable and cost effective transportation, but require a great industry collaboration from Automotive, Transportation. ICT and Cloud.

**Abbreviations**
AV/ADS- Autonomous Vehicle/ Automated Driving System
CV/ITS- Connected Vehicle/Intelligent Transportation System
AD/ADAS- Autonomous Driving/Advanced Driver Assisted System
USDOT- US Department of Transportation
NHTSA- National Highway Traffic Safety Administration
ITS-JPO – ITS- Joint Program Office under NHTSA
FMVSS- Federal Motor Vehicle Safety Standard, US regulatory standard
ISO 26262/SOTIF- International safety standards for product requirements and developments
VII Initiative- Vehicle Infrastructure Integration Initiative
FCC- Federal Communication Commission
UMTRI- University of Michigan Transportation Research Institute
DSRC- Dedicated Short Range Communication
DARPA- Defense Advanced Research Projects Agency
CVPD- Connected Vehicle Pilots Deployment
RSU/OBU- Road Side Unit/On Board Unit
SCMS- Security Credential Management System
SPMD- Safety Pilot Modal Deployment
BSM- Basic Safety Message

# 1. Introduction of CV and AV in US

There are two parallel journeys for CV and AV respectively. Government through POCs, researches and commercial deployments, has funded CV /ITS activities. In Fig. 1, we show three major milestones since 2003. The first milestone occurred during 2003 to 2006, USDOT initiated the VII initiative project for proof of concept work of the ITS for safety and mobility applications. FCC has allocated 75MHz spectrum at 5.9 GHz dedicated for the underlying wireless communication. The second milestone happened during 2011 to 2014, DOT started research pilot program in real-world driving scenarios. During this phase UMTRI conducted a big-scale research pilot project to have installed the safety and mobility applications into 2800 deployed vehicles including cars, trucks etc. in Ann Arbor, Michigan. This research pilot program also helped to publish the first IEEE specification WLAN-based V2X ([IEEE 802.11p](IEEE 802.11p)) in 2012. The specification supports direct communication between vehicles (V2V) and between vehicles and infrastructure (V2I). This technology is referred to as DSRC, which uses the underlying radio communication provided by 802.11p.

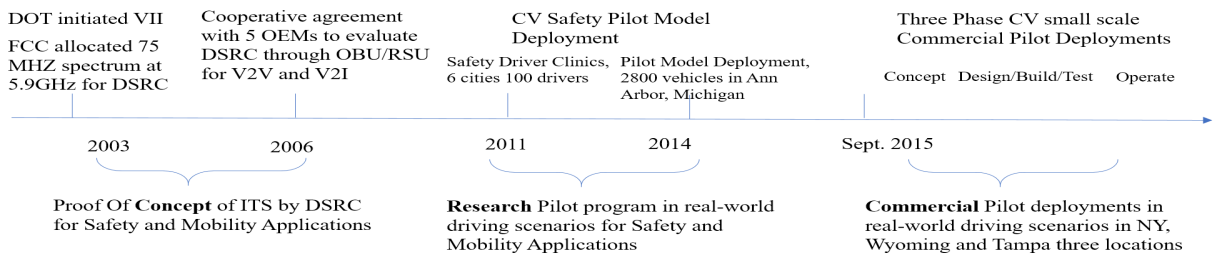

Figure 1- CV/ITS Journey

The third milestone has started since Sept. 2015. ITS-JPO manages a three-phase 5 years program for small-scale commercial deployments in three selected locations including New York, Wyoming and Tampa Florida.

During the third milestone, two major things occur (in Fig. 2). One is that NHTSA cancelled its original proposal of mandating V2V for all the new vehicles starting in 2020, the other was that FCC reallocated the previous 100% DSRC dedicated spectrum into three portions, finally only 14% 75mHz spectrum remaining for DSRC. These are two big setbacks to DSRC community over decade's effort.

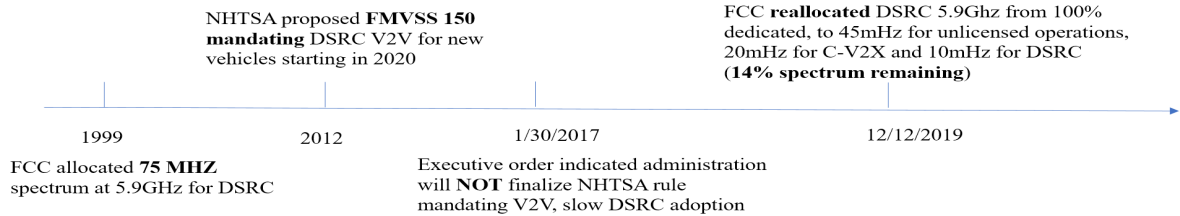

Figure 2- Major Events for DSRC in Regulation and Spectrum

AV/ADS effort could originally track back to the first DARPA sponsored Grand Challenge for driverless car on March 13, 2004. Since then Google started the X's project Chauffeur for AV research. The second wave and golden period for AV came between 2013 and 2017 for the startups. Most of them are located in California bay area. Waymo started small scale commercial Robotaxi service in Dec. 2018 in Phoenix Arizona. Then we saw a few major strategic acquisitions by the Tech companies, for example, recently Intel/Mobileye acquires Moovit for its fleet operation in MaaS market potentially. Amazon acquires Zoox, which the analysts project will help Amazon's delivery service through AV technologies. From those threads, we also emphasize on Tesla's first version of autopilot software that was released in Oct. 2015. Tesla initiates an incremental development for its AV stack by a series of structured ADAS features through continuous software updates. At the end of April, 2020 Tesla passed 4 billion miles on Autopilot through its real-road fleet. People project that Tesla's fleets have at least 1000 times more data than that of any company has in this data-driven domain, where driving data becomes the new "fuel" for AV.

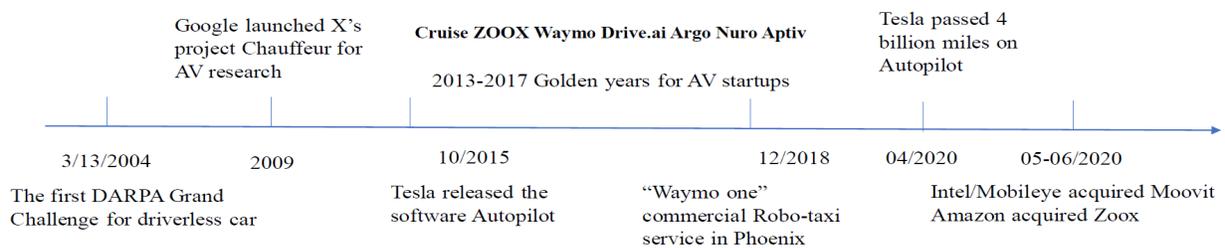

Figure 3- AV from Research to Commercial Deployment

Regarding the policies and regulations for AV/ADS they are still in the early stage and focus on the road testing related state legislations. Federal drives the AV/ADS strategies to facilitate the innovations by removing or exempting any blocking issues for the AV private sector. In US Federal and States have separate responsibilities on vehicles. Federal's role is to formulate the vehicle design and define safety performance standards, and States take the responsibilities of driver registrations, vehicle operations, insurances and liabilities. No major regulations and standards will be enforced in Federal level unless the maturity of AV/ADS reach the right level for potential massive business deployments. Federal has initiated quite a few research projects in this area. One of such projects already comes up with the proposal on new FMVSS standard considerations for vehicles with ADS, it entails a technical translation on the cases of "no human driver" and "no manual steer and brake control".

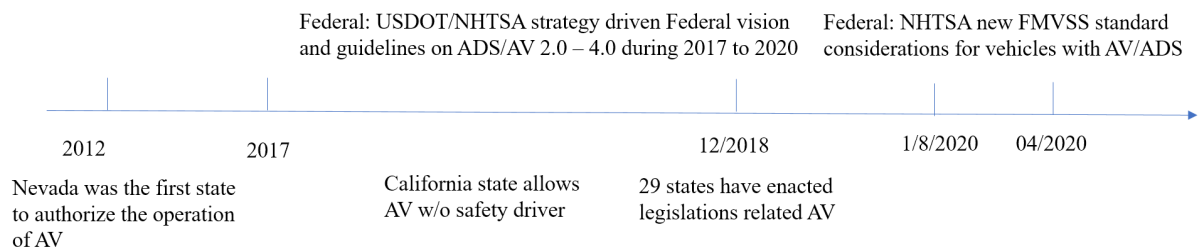

Figure 4- Policy and Regulation Progress in both Federal and State Level

# 2. Safety Pilot Model Developments and Deployments for CV/ITS

### A. Operating Model

In Fig. 5 we show the basic operating model of CV/ITS, which is driven by USDOT through its visions and strategies. There are multiple research areas including technologies, policies and regulations, business models, and State level does the execution of the selected pilot programs. Under each of the pilot programs, the vendors and solution providers from private sector participate the developments and deployments. The independent evaluators have evaluated each pilot program against the project goals.

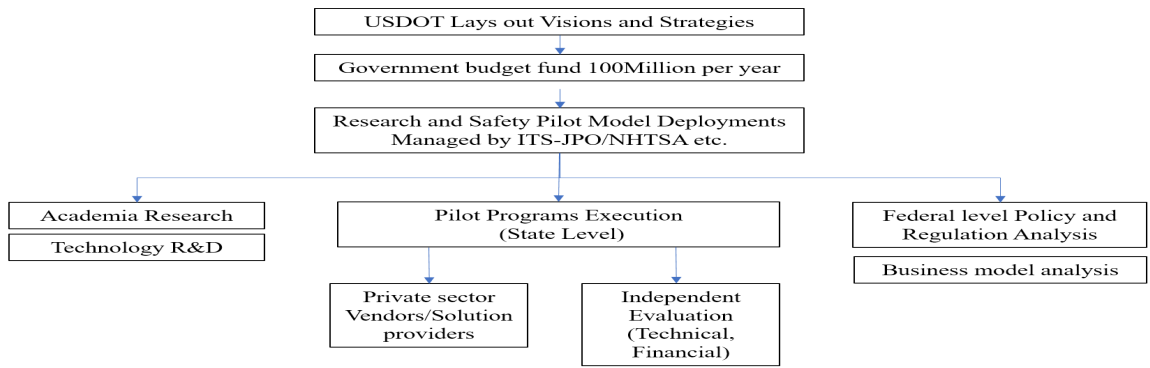

Figure 5- Operating Model of CV/ITS Pilot Program

### B. The Overall Goals of the Connected Vehicle Pilot Deployments

In summary the program goals from USDOT are (refer to Fig. 6):

- To spur early CV **deployments** through wirelessly connected vehicles, mobile devices, infrastructure, and other elements. Data from those multiple sources can be integrated to help make key decisions on safety, mobility and environmental applications.
- To measure the impacts and **benefits** of the real world deployments instead of an isolated or simulated testbed. The activity also includes the differentiating and finding these benefits, identifying what can be attributed to these CV applications and technologies.
- To resolve issues of various real world deployments. People should not only tackle **technical** areas and focus on getting applications to work together, but also put the **institutional** arrangements in place to ensure installation of the technology as well as to manage and govern the sharing of information. Additionally, **financial** arrangements must be made which may integrate the technologies into a financially sustainable model.

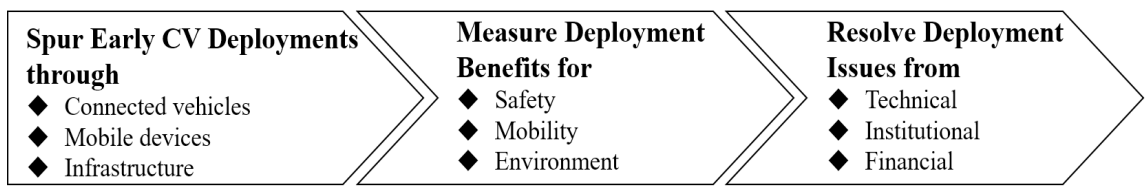

Figure 6- USDOT CVPD Program Goals

## C. CV/ITS System View

In Fig. 7, we draw a typical system functional view of the CVPD programs. It basically includes the in-vehicle devices, RSU, TMC-ITS and external support systems. SCMS is for security and certificate management between vehicles and between vehicle and infrastructure.

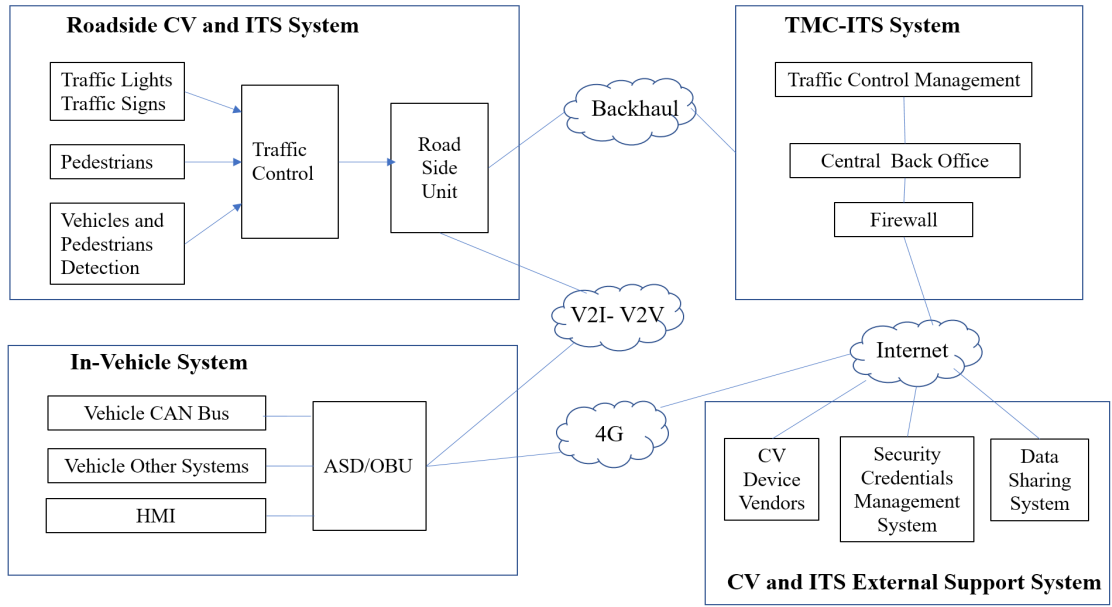

Figure 7- CV System View

## D. Support Scenarios and Applications

NHTSA does a comprehensive analysis on the pre-crashes that could be potentially addressed by V2V technology, V2I technology and combined. Overall, the DOT analysis concluded that, as a primary countermeasure, a fully mature V2X system could potentially address majority of accidents. In Fig.8, we refer to a specific analysis from [1]

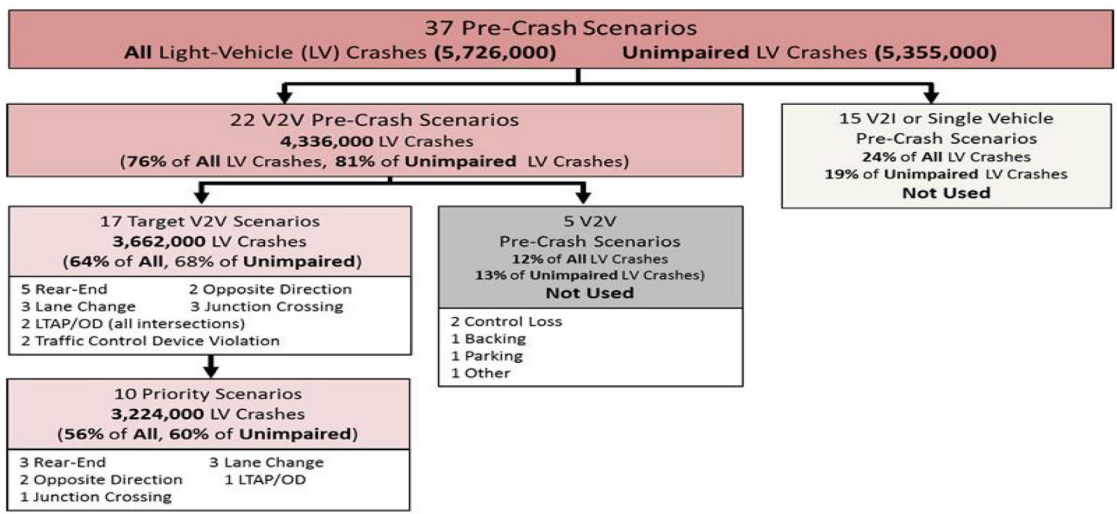

Figure 8- V2V Light-Vehicle Crashes Breakdown from NHSTA Analysis

All the pilot deployments take the above analysis as basic inputs for the design of the scenarios and applications. Although there are many pilot programs, the following table 1 contains the most typical four CVPD projects as reference [1][2][3].

| CVPD | Scale | Scenarios | Applications |
|---|---|---|---|
| Michigan SPMD 2012-2014 | 2800 vehicles, a mix of cars, trucks, and transit vehicles<br>27 RSU covering 75 miles of roadway | V2V, V2I<br>Public streets within a highly concentrated area | . Forward Collision Warning (FCW)<br>. Emergency Electronic Brake Lights (EEBL)<br>. Do Not Pass Warning (DNPW)<br>. Left Turn Assist (LTA)<br>. Intersection Movement Assist (IMA)<br>. Blind Spot Warning & Lane Change Warning (BSW & LCW)<br>. Curve Speed Warning (CSW)<br>. Bridge Height Information<br>. Pedestrian in Signalized Crosswalk Warning (PCW)<br>. Vehicle Turning Right in Front of Bus Warning (VTRW) |
| NYCDOT CVPD 2015- NOW | 4000 ASD/OBU for taxi, buses and cars<br>550 RSUs | V2V, V2I<br>New York City, Manhattan and Brooklyn CBD areas | . VTRW<br>. FCW<br>. EEBL<br>. BSW & LCW<br>. Lane Change Warning/Assist<br>. IMA<br>. Red Light Violation Warning<br>. Speed Compliance<br>. CSW<br>. Speed Compliance/Work Zone<br>. Oversize Vehicle Compliance<br>. Emergency Comms & Evacuation Info<br>. Mobile Ped Signal System<br>. PCW<br>. CV Data for Intelligent Traffic Signal System |
| TAMPA CVPD 2015- NOW | 1000 privately owned vehicles, 8 streetcar trolleys, 10 regional transit buses<br>1000 OBUs, 47 RSUs | 6 use cases: morning backup, wrong way entry, pedestrian safety, transit signal priority, streetcar conflicts, traffic progression, OTA updates, located in the core of Tampa CBD | . End of Ramp Deceleration Warning (ERDW)<br>. EEBL<br>. FCW<br>. IMA<br>. Intelligent Traffic Signal System (I-SIG)<br>. PCW<br>. Transit Signal Priority (TSP)<br>.Vehicle Turning Right in Front of a Transit Vehicle (VTRFTV)<br>. WrongWay Entry (WWE) |
| WYDOT CVPD 2015-NOW | 400 fleet vehicles with OBUs<br>75 RSUs | V2V, V2I Safety in I-80 one of the busiest freight corridors in the region, truck volume is 30-55% of the traffic | . FCW<br>. I2V Situational Awareness<br>. Work Zone Warning (WZW)<br>. Spot Weather Impact Warning (SWIW)<br>. Distress Notification (DN) |

Table 1 Four CVPD Programs Details

Please also refer to [1] for some detailed examples of crash scenarios and vehicle-to-vehicle applications, where rear end collision, lane change and intersection are the three most common crash scenarios in real world car accidents.

There ae three types of aftermarket devices used by those four CVPDs:

- Vehicle Awareness Device: installed by a certified installer and used for BSM transmissions.
- Aftermarket Safety Device: installed by a certified installer for V2V safety applications, receives and transmits BSM, with a driver-vehicle interface.
- Retrofit Safety Device: needs to be connected to the vehicle's data bus for V2V safety applications, receives and transmits BSM, with a driver-vehicle interface.

### E. Assessments, Initial Conclusion and Lessons Learned

In order to assess the readiness for application of vehicle-to-vehicle (V2V) communications USDOT/NHTSA has funded quite a lot researches and evaluations to explore technical, legal, and policy issues relevant to V2V/V2I. Tremendous analysis and research has been conducted thus far, including the technological solutions available for addressing the safety problems identified, the policy implications of those technological solutions, legal authority and legal issues such as liability and privacy. Using those reports and other available information, NHTSA will determine how to proceed with additional activities involving vehicle-to-vehicle (V2V), vehicle-to-infrastructure (V2I), and vehicle-to-pedestrian (V2P) technologies.

The flowing table 2 lists all the dimensions for the assessments [1]:

| Assessment Dimensions | Sub-category | Notes |
|---|---|---|
| 1. Safety need | <ul><li>Crashes potentially addresses by V2V/V2I technology</li><li>Ways of addressing the safety need - scenarios</li><li>Types of V2V/V2I devices</li></ul> | Usefulness |
| 2. Scope and Legal Authority | <ul><li>NHTSA's scope and legal authority in V2V/V2I</li><li>Agency actions within its legal authority</li><li>Non-regulatory actions for stand up communication</li><li>Authority for spectrum for V2V/V2I operations</li></ul> | Authority and regulatory |
| 3. Technical Practicability | <ul><li>Overview of HW and SW components enabling operation</li><li>Interoperability & system limitations</li><li>Global activities and differences in V2V systems</li></ul> | |
| 4. Safety Applications | <ul><li>The safety applications and performance metrics</li><li>Key findings and conclusions for each safety applications</li><li>Driver-vehicle interface</li><li>System compliance and enforcement</li></ul> | |
| 5. Public Acceptance | <ul><li>The importance of public acceptance and potential issues</li></ul> | |
| 6. Privacy Considerations | <ul><li>The importance of privacy considerations and privacy risk assessment</li></ul> | |
| 7. V2V/V2I Communication Security | <ul><li>Security system design and evaluations</li><li>System integrity, management and governance</li></ul> | |

| 8. Legal Liability | <ul><li>Industry liability concerns and solutions, specific concern to SCMS</li><li>Federal liability limiting mechanisms and assessment of industry liability</li><li>NHTSA's assessment of SCMS liability</li></ul> | |
|---|---|---|
| 9. Cost Estimates of V2V/V2I Implementation | <ul><li>Overview of preliminary cost estimates, projected vehicle equipment costs, preliminary system communication costs, security credentials management system cost model</li><li>Economic practicability</li></ul> | |
| 10. Preliminary Effectiveness and Benefits Estimates of V2V/V2I | <ul><li>Analysis of preliminary benefits</li><li>Effectiveness of the safety applications</li><li>Fleet communication rate</li><li>Projected benefits of the technologies</li></ul> | |

Table 2- Assessment Dimensions and Categories

In summary [5], from the assessments conducted by NHTSA and its partners so far, it appears that:

- Basic functions are working, V2V devices installed in light vehicles were able to transmit and receive messages from one another, with a security management system providing trusted and secure communications among the vehicles.
- Safety applications enabled by V2V, examples of which include IMA, FCW, and LTA, have proven effective in mitigating or preventing potential crashes, but additional refinement to the prototype safety applications would be needed before minimum performance standards could be finalized.
- The NHTSA has the legal authority to mandate V2V (DSRC) devices in new light vehicles, and could require them to be installed in commercial vehicles already in use on the road.
- Based on preliminary information, NHTSA estimates that the V2V equipment and supporting communications functions (including a security management system) would cost approximately $341 to $350 per vehicle in 2020.
- The total projected preliminary annual costs of the V2V system fluctuate year after year but generally show a declining trend. The estimated total annual costs range from $0.3 to $2.1 billion in 2020 with the specific costs being dependent upon the technology implementation scenarios and discount rates.
- Regarding safety impacts, NHTSA estimates annually that just two of many possible V2V safety applications, IMA and LTA, would on an annual basis potentially prevent 25,000 to 592,000 crashes, save 49 to 1,083 lives, avoid 11,000 to 270,000 MAIS 1-5 injuries, and reduce 31,000 to 728,000 property-damage-only crashes by the time V2V technology had spread through the entire fleet.

Even with the success of the Safety Pilot Model Deployment in proving that V2V technology can work in a real-world environment on actual roads with regular drivers, additional items need to be in place beyond having the authority to implement a V2V system, in order for a potential V2V system to be successful. These items include wireless spectrum, V2V device certifications, test procedures and performance requirements, communication security, liability, privacy and public acceptance. For more details please refer to [1].

The above issues indicate that through the research conducted to date, NHTSA has a better understanding of the potential of V2V technology, but various aspects of the technology still need further investigation to support transition from a prototype-level to a deployment-level system.

### F. Business Model Analysis

The business of V2X over DSRC is not there yet today. The biggest problem is hard to find a viable business model. However, there are some research and analysis on the potential business models. In Fig. 9, we draw a multi-B2B2C business relation among operator, regulator, OEMs, vehicle owners and insurance companies etc.

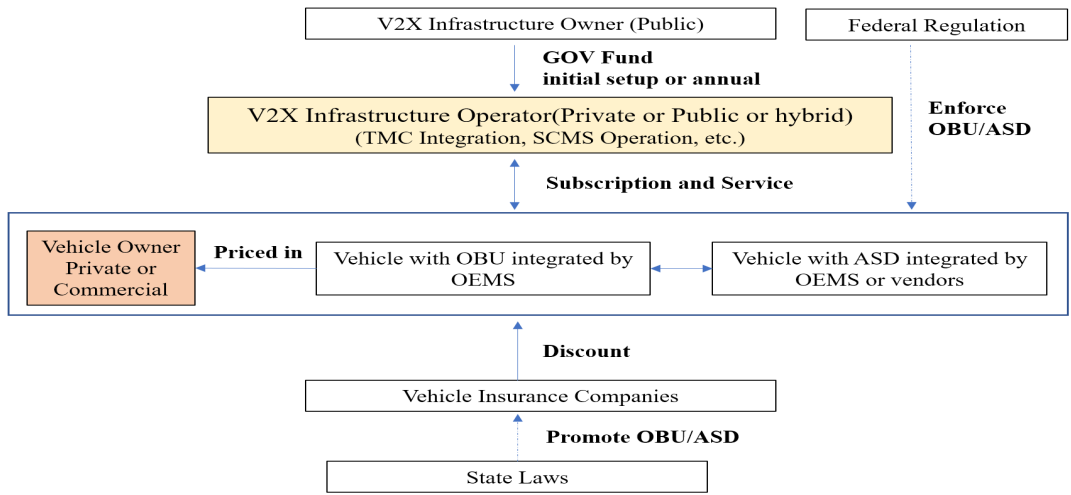

Figure 9- Potential Business Model for V2X over DSRC

A few major factors will help to make this model work:

- The penetration rate of vehicle installation of OBU/ASD will depend on whether federal mandate the need through its regulation. State level law or regulation on insurance may potentially promote the V2X option to the end customers due to the discount on the car insurance fee.
- In order for the V2X operator business maintainable firstly there should have initial investments on infrastructure construction either from government or other channels, secondly the infrastructure operator can financially get the enough service subscriptions from customers for its operation. Another potentially model is the public-owned mode for the infrastructure, similar to the airport operations for airline businesses.
- Innovation field normally initiates through private sectors, it will take much longer period by the government-funded approach. However, especially the initial investment on V2X infrastructure seems to be impossible from any private company. This dilemma has to be resolved in a creative way through the collaborations between public and private sectors.

Note that using C-V2X for CV in US has very limited deployment cases. One of such recent activities leading by QUALCOMM with Virginia DOT is undergoing for case study. With the latest 3GPP R16 for advanced CAV features supporting we can envision there will be more activities in CAV use cases' testing and verification.

## 3. AV Journey Driven by Market from Private Sector

AV has been a hot topic for decades. There are a lot of information in this area from researches to industry practices. In this paper, we are not going to emphasize on the analysis of different approaches, technologies and related business models. Finally, we come up with some recommendations for CAV developments and deployments from industry perspectives.

### A. Approaches and Journeys to AV

In Fig. 11, we draw a comparison between two camps of players. One is from OEMs where Tesla is a typical disruptive OEM to make the software-defined vehicle, and Volkswagen, the traditional OEM, takes a digital transformation journey aiming to be a software vehicle company. The second camp is from AV tech companies, where Waymo and Cruise are two typical AV startups, Nividia and Intel are playing the platform strategies from bottom up (Though we see a trend that the leading platform company are starting to operate the feet in the new MaaS business market). We have done some high level analysis for these two camps in strategies, technologies, business models and related markets. Obviously traditional OEMs are facing huge challenges in both in-house software capabilities and new digital and intelligent technologies for AV. In the other side, tech companies have solid reservations in software and technologies, but lacks of vehicle domain knowledge and operation experience. Tesla has created a disruptive way to the software-defined vehicle with a much simpler EV architecture. With the centralized computing E/E architecture Tesla can greatly simplifies its supply chain into a more IT style iterative development and production environment.

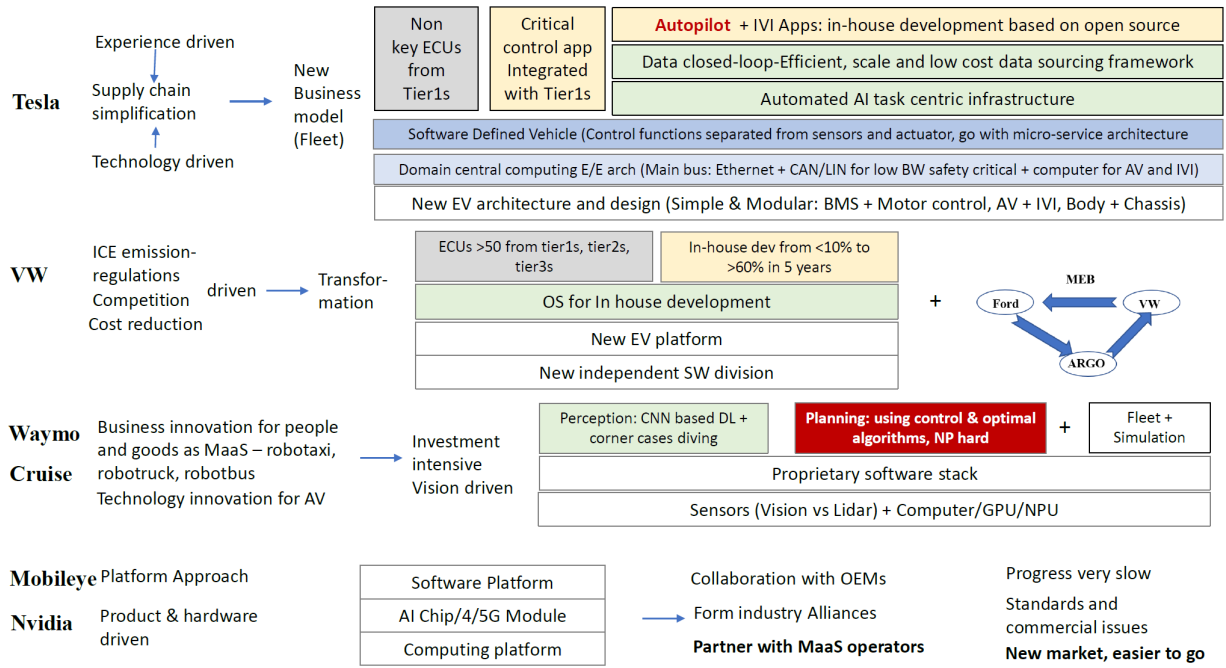

Figure 10- AV Approaches by Different Players

## B. Technologies and business models analysis on AV

We also analyze the strategies, technologies and business models for different players in table-4.

| Players | Strategies | Technologies | Market & Business Models |
|---|---|---|---|
| Tesla | ➢ From Autopilot to Full Self Driving, incremental ADAS supports for one AV vision<br>➢ From private ownership car to fleet operations | . Cost effective sensors, camera based for better scalable solution for all driving environments, data-driven incremental development | . Ownership car<br>. Fleet for MaaS |
| Traditional OEMs Volkswagen | ➢ Silo ADAS functions from different vendors<br>➢ Separate AV strategies<br>■ From in-house dev to investment on AV tech | . Source AV technologies from startups or tech companies | . Ownership car<br>. Value added service in MaaS |
| Waymo | ➢ Dev general software and algorithms for all scenarios in L4+<br>➢ Start to tackle easier scenario for feasible business deployments in Robotaxi and Robotrunk | . Lidar/map based with high cost technology driven approach<br>. Invest in sensor hardware<br>. Need customization from one site to another for deployments | . MaaS for people and goods<br>. Licensing AV software to OEMs (long term) |
| Other AV startups | ➢ Passed the cash-burning phase, strive for survival in niche or simpler scenarios for business | . Most are taking Lidar/map based approach<br>. Some are taking camera-based approach | . MaaS for goods in niche markets |

| | | deployments in trunk, logistics & delivery, bus etc. | | |
|---|---|---|---|---|
| Tech Platform players<br>Nivdia<br>Intel/Mobileye | ➢ | Start with hw platform, and incrementally build software ecosystem for AV in perception area | . AI chips<br>. Build software enabling platform for AV applications | . Platform SPs for ownership and MaaS<br>. Explore the role as fleet SP directly (Mobileye) |
| | ➢ | Look for operations and data for fleet management through new MaaS market | | |
| Commerce Platform Player<br>Amazon | ➢ | Has enough business cases and values for AV in logistics and goods deliveries | . Acquired technologies directly for its business scenarios | . AV technologies enables commerce core business for cost reduction and efficiency |
| | ➢ | From investment to direct technology acquisition (Zoox etc.) | | |

Table 3- Analysis on Technologies and Business Models

Besides the information in Table 3, there are a few extra things to mention:

- The OEM-to-OEM alliance in AV seems to be going away, replaced by OEM(S)-to-Tech Company. One specific example is the recent breakup between BMW and Daimler's partnership in AV joint venture.
- Business value drives the technology acquisitions. Amazon's 1.3 billion acquisition of Zoox follows this trend, where billions of business values exist as long as AV technology can take off in those relatively fixed and simpler driving environments.
- In order to be product ready the AV modules including both hardware and software need to integrate with the E/E architecture of the vehicle. Due to the performance requirements of AV in computing, real-time processing in both low latency and high bandwidth throughput cases, and service agility demands for iterative data-driven and development cycle etc. Without E/E architecture upgrade there is no way for the traditional vehicle to fit the AV needs by meeting all those requirements. OEMs or new EV startups have to find a transformation route for their vehicle architecture to be a more native development and product environment for AV.

**C. Latest Effort from CAV**

USDOT has initiated CV effort for more than a decade, the developments and deployments have mainly been targeting to provide safety and mobility applications for car, truck and bus drivers. It is a similar approach as OEMs does for their ADAS applications. Instead, applying the connected information and intelligence to the AV domain (so-called CAV) is still one open research topic today.

The following Table 4 list three exiting research approaches [12] and one recommendation by this paper. We also do a high-level analysis on the feasibilities of each approach from both technology and business views. We estimate that in the next 3 to 5 years CAV area will attract intensive research and development activities and investments. The connected information and intelligence have to be deeply involved into the AV context to provide critical information for vehicle seeing more and farther range for the perception, and help collecting the driving intensions of other road participants for the AV path planning.

| Approaches | For AV Perception Fusion | For AV Planning Fusion | Technology and Business Analysis |
|---|---|---|---|
| Low level- sensors raw data sharing through V2V & V2I | . Sensors' data sharing among vehicles<br>. Pros- no information loss<br>. Cons- large amount of data communication | | . From business perspective for vehicle from different OEMs or AV companies the perception is competitive capability, no companies will be willing to share such sensors' data |
| High level- object detection results sharing through V2V & V2I | . Object detection results sharing among vehicles<br>. Pros- reduced amount of data for communication<br>. Cons- some detailed information will be lost | | The same business argument as above |
| Intermediate level- Feature level information sharing | . Neural network Feature level information sharing among vehicles<br>. Right amount data and information | | . From technology perspective it's not so feasible since different vendors may have different neural network topologies<br>. Business perspective- the same argument as above |
| Safety critical non-competitive information sharing, treated as one modified DSRC approach, but more tied to the AV context for both perception and planning | Some examples:<br>. The basic [P, S,V, A, Yaw] of each dynamic objects in V2V, V2I and V2P scenarios<br>. Traffic sign/lights etc. | Some examples:<br>. The vehicle intension of turning, braking, accelerating etc. for path planning<br>. Traffic flows information for route planning | . Sharing critical non-competitive information and data is feasible from both technology and business views<br>. Can leverage a lot of real-world deployment experience from USDOT DSRC approach for safety and mobility applications |

Table 4- Different Research Comparison on CAV

CAV should also leverage the scenarios and applications that DSRC has done so far. More specifically, as shown in Figure 11 [6]:
- Reuse of DSRC/C-ITS established service and app layers and do the safety application standard extension if necessary.
- Reuse exiting security and transportation layer if applicable（Defined by ISO, ETSI, and IEEE 1609 family）.

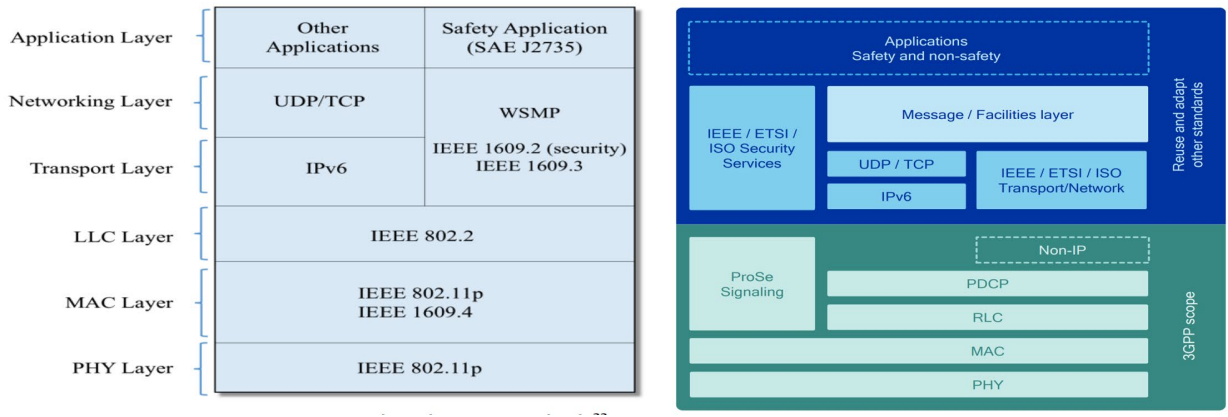

Figure 11- DSRC/C-ITS Stack vs C-V2X Stack

CV involves multi-industry collaborations from Automotive, Transportation, ICT and Clouds. Not only does it needs standard communication protocols, but also wants format specifications on the exchanging information models and supporting cooperative information framework, thus to facilitate collaborations and conquer the barriers of information flows.

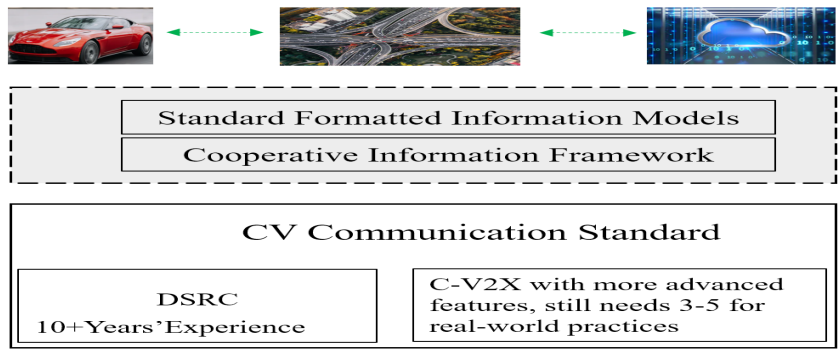

Figure 12- CV through Vehicle, Road and Cloud Collaborations

## 4. Recommendations for CAV from Industry Perspective

It's very challenging for AV commercial deployments in real world driving scenarios, and those promises have been pushed out for multiple times. We believe that the cooperative CAV may have a better chance for a viable commercial deployment. However, a few fundamental decision points and tasks need to make and execute before turning the dream into a reality. Those include the construction of CAV infrastructures, the architecture upgrade for software-defined mobility, key technologies advances in algorithms and new toolsets, regulation enablers and new business drivers. In Fig. 13, we design a high-level approach with recommendations for each layer. Some of efforts need open industry collaborations. We envision that CAV can come into the reality much earlier than people think it would be only when those collaborations are on the horizon.

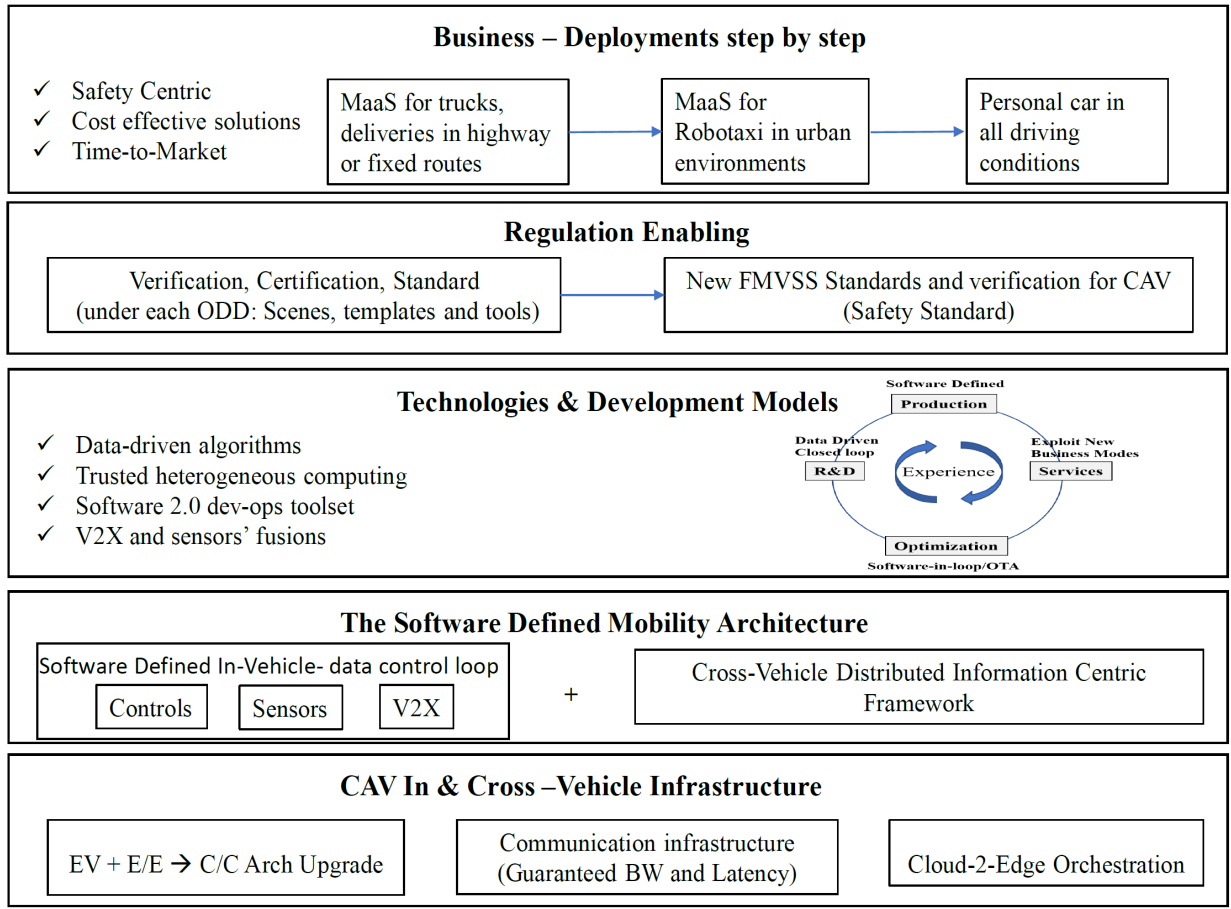

Figure 13- Industry Collaboration Recommendations

We would like to emphasize on a few critical tasks:

- Business and technology-driven: It's a both top-down and bottom-up approach, for example, viable business builds on the CAV infrastructure investment, solid architecture and technologies will make the cost effective business possible.
- Infrastructure investment and viable business model: The biggest problem of DSRC trials is that it has not established a viable business model. It is not just an issue lack of regulation or government fund; it has not found a self-maintainable operation model yet. CAV infrastructure operators have to build the "differentiated values" for the businesses it supports.
- New markets: The AV cash-burning period has already passed. CAV business deployments needs to be more realistic. In cities most people and commercial companies cares more about a safe, efficient and cost effective mobility transportation. CAV approaches can stage as phases with easier and valuable scenarios as the starting point. The three-phase approach is to start with trucks, deliveries and logistic market in highway and fixed driving routes. Then, move to a more challenging urban driving environment, Robotaxi. The third phase will be the ownership car that people can drive anywhere they want for all the driving scenarios.
- Architecture and technologies: We have to realize that the 100+ years' old automotive industry is the mechanical and electronic hardware centric business. The pace of automobile can nerve match new business requirements or the pace of latest ICT technologies without any fundamental vehicle architecture upgrade. The embedded system architecture with hardware and function centric is blocking the realization of the intelligence and digitalization for the vehicle. The upgrade of in-vehicle architecture will enable a trusted-control loop for AV safety applications. The cross-vehicle framework will support a seamless information centric architecture for connected information and intelligence. For technical details please refer to author's another IEEE paper [13]. We also need to realize that digitization and intelligence is driving a new iterative development model, where features will be delivered through software in weeks instead of years.

- Regulation and policy [10]: Without them, there will not be any commercial services. However, they need solid supports from standards, best practices, toolchains for testing, verification and certification. Guidelines and exemptions are to encourage the innovations in the incubation period. Formalized methodologies and verification tools under each ODD (Operational Design Domain) will greatly help to drive the new safety standards, regulations and verification processes to evolve. We believe that CAV regulations should be enacted gradually by each well-defined ODD (including driving scenes, road type, vehicle velocity and dynamics, weather, illumination etc.) for CAV with solid technologies.